\documentclass[preprint,showpacs,preprintnumbers,amsmath,amssymb]{revtex4}
\usepackage{CJK}

\usepackage{graphicx}
\usepackage{bm}

\begin{document}
\begin{CJK*}{GBK}{song}
\preprint{USTC-ICTS-11-03}

\title{ Entropy Bound Derived from the New Thermodynamics on  Holographic Screen}
\author{Wu Guang}
\email{rocky29@mail.ustc.edu.cn} \affiliation{Department of Modern Physics, University of Science and Technology of China, Hefei 230026, China}

\author{Gu Wei}
\email{guwei@mail.ustc.edu.cn} \affiliation{Interdisciplinary
Center for Theoretical Study, University of Science and Technology
of China, Hefei 230026, China} \affiliation{Department of Modern Physics, University of Science and Technology of China, Hefei 230026, China}

\begin{abstract}

We introduce a new interpretation of chemical potential and show that
holographic entropy is entropy bound, which is supported by two ideal cases
discussed in detail. One is sparse but incompressible liquid like a star of
uniform density and the other is a screen at infinity in spherically
symmetric sapcetime. Our work is based on the new scenario of entropy force
and holographic thermodynamics, and the Brown-York quasi-local
energy.\bigskip
\end{abstract}

\maketitle
\section{\protect\bigskip Introduction}

In 1970s, the discovery of black hole entropy, the four laws of classical
black hole mechanics and Hawking radiation revealed a mysterious but
charming connection between gravitation and thermodynamics. In 1995,
Jacobson derived the Einstein equations as equilibrium equation of state by
assuming proportionality of entropy and horizon area, and the Clausius
relation. Since then, a large number of papers have sprung up to generalize
the approach to other gravity theories \cite{Eling:2006aw}\cite{Elizalde and
Silva}\cite{Brustein:2009hy}. Last year, Verlinde \cite{Verlinde:2010hp}
argued the gravity is not only an emergent force, but particularly an
entropy force. His work is motivated by the Jacobson's work as well as the
gauge/gravity correspondence.

Recently, Li and his collaborators gave a new scenario of entropy force and
holographic thermodynamics \cite{Gu:2010wv}. First, a surface stress tensor
is introduced on a holographic screen while Brown-York quasi-local energy
\cite{Brown} is adopted to replace Verlinde energy. This tensor would give
both the surface energy and pressure. Then, a closed screen used by Verlinde
is extended to an open or closed one. Finally, a time-like screen is used
rather than a null screen. The thermodynamics is self-consistent and back to
the Jacobson's form while the time-like screen is approaching the null
screen.

As an important part, chemical potential on holographic screen is discussed
in detail, and specific examples are calculated in Li's paper. For example,
a screen sweeping vacuum outside a black hole horizon gives a form of
chemical potential according to thermodynamic relations.

In this paper, we propose a different interpretation of chemical potential
on holographic screen. We ponder that such holographic entropy is the
entropy bound, independent of concrete matter, since the entropy is given
within an ideal system. There are two reasons in favor of our point. One is
the Verlinde temperature is general, having nothing to do with the
composition of matter. It only requires the information of metric. The other
is Birkhoff's theorem ensure the irrelevance between exterior metric and
interior matter composition.\qquad

In section III, we discuss two ideal cases, each of which will lead to the
entropy bound respectively. For the first case, we analyze the system with
sparse but incompressible liquid like star of uniform density and then give
a holographic derivation of Bekenstein entropy bound. For the second case,
we consider a spherically symmetric system, and compute its holographic
entropy on a screen at infinity. Since the temperature we use is not related
to specific matter system, the entropy we obtain is not the usual
statistical one (like the black hole entropy). At infinity, we also can not
recognize the form of system, so the holographic entropy which we calculated is the entropy
bound. In this paper, we choose the natural units(ie. $\hbar=c=1$)

\section{Preliminary}

In this paper, we need the form of the surface stress tensor and other
indispensable quantities, thus we have to review \cite{Gu:2010wv}\cite%
{Miao:2011er}, and lay the foundation of our calculation in section III.

Introducing some basic definitions is beneficial. We set a 2+1 dimensional
timelike hypersurface $\Sigma $ embedded in the 3+1 dimensional spacetime $M$%
. $x^{a}$, $g_{ab}$, $\nabla _{a}$, $y_{i}$, $\gamma _{ij}$ and $D_{i}$
(here $a$,$b$ run from 0 to 3, and $i$,$j$ run from 0 to 2) denote the
coordinates, metric and covariant derivatives of $M$ and $\Sigma $
respectively. The hypersurface $\Sigma $ can be described by a scalar
function
\begin{equation}
f_{\Sigma }(x^{a})=c.  \label{fSigma}
\end{equation}

Analogous to Jacobson's procedure \cite{Jacobson}, we suppose an energy flux
$\delta E$ passing through an open but time-like patch on holographic screen
$d\Sigma =dAdt$
\begin{equation}
\delta E=\int_{\Sigma }T_{ab}\xi ^{a}N^{b}dAdt,  \label{E1}
\end{equation}%
where $T_{ab}$ is the stress tensor in the bulk $M$, $\xi ^{a}$ is the
Killing vector, and $N^{a}$ is the unit vector normal to $\Sigma $.
Obviously, a vector normal to $\Sigma $ is proportional to $g^{ab}\nabla
_{b}f$, and can be normalized as
\begin{equation}
N^{a}=\frac{g^{ab}\nabla _{b}f_{\Sigma }}{\sqrt{\nabla _{b}f_{\Sigma
}\nabla ^{b}f_{\Sigma }}}.  \label{Na}
\end{equation}

From the viewpoint of the holographic principle, every physical process in
the bulk is corresponding to a process on the holographic screen. When an
energy flux $\delta E$ passes through the screen $\Sigma $ from the left
side to the right side, the energy will change on the right side, and thus a
corresponding change on the screen due to the holographic principle. The
energy density $\varepsilon $ and energy flux $j$ on $\Sigma $ take the form
\begin{equation}
\varepsilon =u_{i}\tau ^{ij}\xi _{j}\ ,\ j=-m_{i}\tau ^{ij}\xi _{j},
\end{equation}%
where $\tau ^{ij}$ is the energy-momentum tensor on $\Sigma $, $u_{i}$ and $%
m_{i}$ are the unit vectors normal to the screen's boundary $\partial \Sigma
$ (we will give the expressions of $u_{i}$ and $m_{i}$ below Eq.(\ref{Mi})),
and $\xi _{i}$ is the Killing vector on $\Sigma $. Naturally, $\tau ^{ij}$
should only rest on the geometry of the boundary, thus we suppose (in \cite%
{Miao:2011er}, a more general and appropriate ansatz has been given, which,
however, do not influence our discussion)
\begin{equation}
\tau ^{ij}=n(K^{ij}-K\gamma ^{ij}),  \label{t}
\end{equation}%
where $n$ is a constant to be determined, and $K^{ij}$ is the extrinsic
curvature on $\Sigma $ defined as
\begin{equation}
K_{ij}=-e_{i}^{a}e_{j}^{b}\nabla _{a}N_{b}.
\end{equation}%
Here $e_{i}^{a}=\frac{\partial x^{a}}{\partial y^{i}}$ is the projection
operator satisfying $N_{a}e_{i}^{a}=0$. It should be stressed that Eq.(\ref%
{t}) is the key ansatz in this scheme to derive the Einstein equations. The
total energy variation on the screen then becomes
\begin{equation}
\delta E=(u_{i}\tau ^{ij}\xi _{j})dA|_{t}^{t+dt}-m_{i}\tau ^{ij}\xi
_{j}dydt=-\int_{\partial \Sigma }(M_{i})\tau ^{ij}\xi _{j}\sqrt{h}%
dz^{2}=-\int_{\Sigma }D_{i}\tau ^{ij}\xi _{j}dAdt.  \label{EE2}
\end{equation}%
Notice that there are two contributions to $\delta E$, namely the variation
of the energy density $\rho $ and the energy flux $j$ flowing into the
screen. $h$ is the determination of the reduced metric on $\partial \Sigma $%
, with $D_{i}$ the covariant derivative on $\Sigma $ and $M_{i}$ a unit
vector in $\Sigma $ and normal to $\partial \Sigma $. Let us choose a
suitable function $f_{\partial \Sigma }(y^{i})=c$ on $\Sigma $ to denote the
boundary $\partial \Sigma $, then we have
\begin{equation}
M^{i}=\frac{\gamma ^{ij}D_{j}f_{\partial \Sigma
}}{\sqrt{D_{j}f_{\partial \Sigma }D^{j}f_{\partial \Sigma }}}.
\label{Mi}
\end{equation}%
When $M^{i}$ is along the direction of $dy^{0}$($dt$) it becomes $u^{i}$,
and when\ it is along the direction of $dy^{i}$($dy^{1},dy^{2}$) it becomes $%
m^{i}$.

From Eq.(\ref{t}) and the Gauss-Codazzi equation $%
(R_{ab}-Rg_{ab}/2)N^{a}e_{i}^{b}=-D_{j}(K^{ij}-K\gamma ^{ij})$, one can
rewrite the Eq.(\ref{EE2}) as%
\begin{equation}
\delta E=\int_{\Sigma }m(R_{ab}-\frac{R}{2}g_{ab})\xi
^{i}e_{i}^{a}N^{b}dAdt. \label{e22}
\end{equation}%
In the following derivations, we will only focus on the quasi-static process
for simplicity. Recalling that we have use Eq.(\ref{fSigma}) to denote the
holographic screen $\Sigma $. In the quasi-static limit, $f_{\Sigma }(x^{a})$
is independent of time, so $N_{a}\sim (0,\partial _{1}f_{\Sigma },\partial
_{2}f_{\Sigma },\partial _{3}f_{\Sigma })$. Note that $\xi ^{a}\simeq
(1,0,0,0)$ in the quasi-static limit, thus we have $N_{a}\xi ^{a}\rightarrow
0$. Then the Killing vector $\xi _{i}$ on $\Sigma $ can be induced from the
Killing vector $\xi _{a}$ in the bulk $B$,
\begin{equation}
\xi _{i}=\xi _{a}e_{i}^{a},\ \ D_{(i}\xi _{j)}=\nabla _{a}(\xi
_{b}-N_{c}\xi ^{c}N_{b})e_{(i}^{a}e_{j)}^{b}=K_{ij}N_{a}\xi
^{a}\rightarrow 0.
\end{equation}%
Thus, in the quasi-static limit, we can rewrite Eq.(\ref{E1}) as
\begin{equation}
\delta E=\int_{\Sigma }T_{ab}\xi ^{i}e_{i}^{a}N^{b}dAdt.  \label{E11}
\end{equation}%
From the Eq.(\ref{E11}) and Eq.(\ref{e22}), it follows
\begin{equation}
n(R_{ab}-\frac{R}{2}g_{ab})e_{i}^{a}N^{b}=T_{ab}e_{i}^{a}N^{b}.
\label{Einstein}
\end{equation}%
Note that $g_{ab}e_{i}^{a}N^{b}=0$, thus we can add a cosmological constant
term $\Lambda g_{ab}$ to the left of the above equation. Defining the
Newton's constant as $G=\frac{1}{8\pi n}$, we then get the Einstein
equations,
\begin{equation}
R_{ab}-\frac{R}{2}g_{ab}+\Lambda g_{ab}=8\pi GT_{ab}.  \label{Einstein1}
\end{equation}%
Substitute $n=\frac{1}{8\pi G}$ into Eq.(\ref{t}), we obtain,
\begin{equation}
\tau ^{ij}=\frac{1}{8\pi G}(K^{ij}-K\gamma ^{ij}),
\end{equation}%
which is just the quasi-local stress tensor defined in \cite{Brown}. As
emphasized in \cite{Brown}, $\tau ^{ij}$ characterizes the entire system,
including contributions from both the gravitational field and the matter
fields. In addition, generally, $\tau ^{ij}$ need to be subtracted in order
to derive the correct energy.

To find the relationships between the energy in the bulk and on the
holographic screen, we require a closed screen. Using Eq.(\ref{E1}) and Eq.(%
\ref{EE2}), together with the conservation of energy in the bulk%
\begin{equation}
\int \sqrt{-g}\nabla _{a}(T^{ab}\xi _{b})dx^{4}=-(U_{a}T^{ab}\xi
_{b})dV|_{t}^{t+dt}+\int_{\Sigma }T_{ab}\xi ^{a}N^{b}dAdt=0\text{,}
\label{energy}
\end{equation}%
we get%
\begin{equation}
(u_{i}\tau ^{ij}\xi _{j})dA|_{t}^{t+dt}=(U_{a}T^{ab}\xi _{b})dV|_{t}^{t+dt},
\end{equation}%
where the fact that a closed screen has no time-like boundary has been used.
This shows the relationship between the bulk energy and the energy on the
holographic screen, which also make the logic in section III complete.
Obviously, they differ from each other by a time independent constant at
most. This constant is associated with the subtraction scheme of the energy
on the screen, and generally it depends on the position of the screen. Due
to the constant independent of time, the subtraction scheme does not affect
our derivation of the Einstein equations.

\section{Calculation}

In this section, we follow the procedure in Sec. VI of \cite{Brown}, and
delve into the first law of thermodynamics on the holographic screen.

\bigskip Let us consider a 2+1 dimensional holographic screen $\Sigma $ with
a constant time slice $B$. For simplicity, we will deal with the case of a
static, spherically symmetric spacetime%
\begin{equation}
ds^{2}=-N^{2}dt^{2}+h^{2}dr^{2}+r^{2}(d\theta ^{2}+\sin ^{2}\theta d\phi
^{2})  \label{metric}
\end{equation}%
where $N$ and $h$ are functions of $r$ only, and $\Sigma $ is a hypersurface
described by $r=R=constant.$ Mass is distributed around the origin point
with spherical symmetry. We use the quasi-local energy in \cite{Brown} to
investigate the first law of thermodynamics on screen.

As mentioned in Sec. II, an appropriate subtraction scheme for the stress
tensor on $\Sigma $ is necessary to derive the correct energy. For the
asymptotically flat space, Brown and York propose to deduct the
contributions of energy from the flat background spacetime, which gives rise
to the same result at infinity as the ADM energy for asymptotically flat
space. Thus, we concentrate on spherically symmetric asymptotically flat
spacetime in this section. More details of the subtraction scheme available
from Sec. VI of \cite{Brown}, we merely use the results without proof.

We choose Newton constant $G=1$ for simplicity. Following the subtraction scheme used in \cite{Brown}, the proper surface
energy density on screen becomes
\begin{equation}
\varepsilon =u_{i}\xi _{j}\tau ^{ij}=\frac{N}{4\pi }(\frac{1}{r}-\frac{1}{rh%
})|_{R},
\end{equation}%
while the proper surface momentum density $j_{a}$ vanishes due to spherical
symmetry. And the pressure $p$, trace of the spatial stress, is
\begin{equation}
p=\frac{1}{2}\sigma _{\dot{a}\dot{b}}\tau ^{\dot{a}\dot{b}}=\frac{1}{8\pi }(%
\frac{N^{\prime }}{Nh}+\frac{1}{rh}-\frac{1}{r})|_{R}.  \label{p}
\end{equation}%
$u_{i}$ is the unit vector normal to boundary $\partial \Sigma $, and $\xi
_{i}\ $is the Killing vector on $\Sigma $.$\ \sigma _{\dot{a}\dot{b}}$ is
the induced metric on two-boundary $B$. Both $%
\varepsilon $ and $p$ are measured by observer at infinity and vanish in
flat spacetime. Thus, the quasi-local energy on a time slice $B$ is%
\begin{equation}
E=\int_{B}d^{2}x\sqrt{\sigma }\varepsilon =N{(r-\frac{r}{h})}|_{R}.
\label{Energy}
\end{equation}%
Introduce the Verlinde temperature on the screen as%
\begin{equation}
T=\frac{1}{2\pi }e^{\phi }N^{b}\nabla _{b}\phi =\frac{1}{2\pi }\frac{%
N^{\prime }}{h}.
\end{equation}
Then, we assume that the full form of the first law of thermodynamics on
screen is
\begin{equation}
dE=TdS-pdA+\mu dN_{f},  \label{law1}
\end{equation}%
where $S$ is the entropy of screen, $\mu $ is the chemical potential and $%
N_{f}$ is the number of freedom. Here we suppose $Nf=A$ as in Verlinde's paper.

We will discuss an ideal model in which mass density $\rho $ is a
sufficiently small constant independent of $r$. It suggests incompressible
liquid and infinite sound speed like stars of uniform density. Thus, the
holographic entropy we calculate is rather the entropy bound of ordinary
matter. It will give some intuitions about specific physical system. In GR,
with a constant $\rho $, there are exact solutions of metric \cite{sean}:
\begin{equation}
N(r)=\frac{3}{2}(1-\frac{2M}{R})^{\frac{1}{2}}-\frac{1}{2}(1-\frac{2Mr^{2}%
}{R^{3}})^{\frac{1}{2}}=\frac{3}{2}(1-\frac{2M}{R})^{\frac{1}{2}}-\frac{1}{2%
}(1-\frac{8\pi \rho r^{2}}{3})^{\frac{1}{2},}
\end{equation}%
\begin{equation}
h(r)=\frac{1}{\sqrt{1-2\frac{m(r)}{r}}}=\frac{1}{\sqrt{1-\frac{8\pi \rho
r^{2}}{3}}},
\end{equation}%
where $R$ is the radius of matter distribution and $M$ is the total mass.

We initially suspend the chemical potential and assume the first law of
thermodynamics on the holographic screen
\begin{equation}
dE=TdS-pdA.  \label{no che}
\end{equation}%
Using Brown and York's energy definition, relative quantities are as
follows:
\begin{equation}
T=\frac{N^{\prime }}{2\pi h}=\frac{2\rho r}{3},
\end{equation}
\begin{equation}
p=\frac{1}{2}\sigma _{\dot{a}\dot{b}}\tau ^{\dot{a}\dot{b}}=\frac{1}{8\pi}(%
\frac{N^{\prime }}{Nh}+\frac{1}{rh}-\frac{1}{r})=\frac{\rho }{4}(1-\sqrt{1-%
\frac{2M}{R}})-\frac{\pi \rho ^{2}}{18}(1+3\sqrt{1-\frac{2M}{R}}),
\end{equation}

\begin{equation}
E=\frac{4\pi \rho r^{3}}{3}(\frac{3}{2}(1-\frac{2M}{R})^{\frac{1}{2}}+\frac{%
2\pi \rho r^{2}}{3}-\frac{1}{2}),
\end{equation}%
thus we can get
\begin{equation}
S={3\pi \sqrt{1-\frac{2M}{R}}r^{2}}+\frac{2\pi ^{2}\rho r^{4}}{3}%
(1+3\sqrt{1-\frac{2M}{R}})={3\pi \sqrt{1-\frac{2M}{R}}r^{2}}+\frac{%
8\pi ^{2}\rho r^{4}}{3}+o(\rho ^{2})\text{.}
\end{equation}%
When $\rho \Longrightarrow 0$, we have $S=3\pi r^{2}$. Actually,
without matter, no normal entropy appears. Thus, we introduce a holographic
chemical potential, and then the 1st law is back to the form of Eq.(\ref{law1}%
). Then, we immediately have
\begin{equation}
\mu =\frac{\sqrt{1-\frac{2M}{R}}\rho r}{2}=\frac{\rho r}{2}+o(\rho ^{2})=%
\frac{\rho r}{2}=\frac{3T}{4}.  \label{pp}
\end{equation}%
This is proportional to $T$ and $\rho r$, which denotes the influnce of particles passing through the screen.
Finally, we get the entropy bound $S=\frac{%
8\pi ^{2}\rho r^{4}}{3}\approx2\pi Er$ rather than the entropy of concrete matter.

As the second model, we place a screen sweeping vacuum at infinity in the spherically symmetric space.
According to Birkhoff's theorem, no matter what the composition of matter is
and how it distributes (up to spherical symmetry) and evolves, the geometry of
outer space-time is described by Schwarzschild vacuum solution, such as
black hole, star system. This is very similar to the case with constant bulk
density. Again, the specific mathematical calculation of entropy is just
entropy bound of some special physical system.

Still without the chemical potential like Eq.(\ref{no che}), those quantities
have different forms:%
\begin{equation}
T=\frac{N^{\prime }}{2\pi h}=\frac{M}{2\pi r^{2}},
\end{equation}

\begin{equation}
p=\frac{1}{2}\sigma _{\dot{a}\dot{b}}\tau ^{\dot{a}\dot{b}}=\frac{(1-\sqrt{1-%
\frac{2M}{r}})r-M}{8\pi r^{2}},
\end{equation}

\begin{equation}
E=Nr(1-\frac{1}{h})=2M+r(\sqrt{1-\frac{2M}{r}}-1).
\end{equation}%
The entropy is approximately%
\begin{equation}
S\approx {\pi (-r^{2}+\sqrt{1-\frac{2M}{r}}r(3M+r)+3M^{2}\log 2(-M+r+%
\sqrt{1-\frac{2M}{r}}r))}.
\end{equation}%
When $r\Longrightarrow \infty $, the mass $M$ can be treat as a small
quantity, then we expand $S$ as%
\begin{equation}
S\approx {2\pi rM+(\frac{\pi }{2}-3\pi \log (M)+3\pi \log (2r))M^{2}-%
\frac{(5\pi M^{3})}{r}}\approx {2\pi Mr}.
\end{equation}%
This is the Bekenstein bound. The entropy $S\Longrightarrow 0$ while the ADM
mass $M\Longrightarrow 0$, therefor the holographic chemical potential
vanishes.

The reason why we lay the screen at infinity is attributed to the non-locality of gravity.
It requires an asymptotic flat spacetime to define the total energy.
To include the whole physics, especially the gravity, we choose an isolated system which have
infinit boundary to satisfy the asymptotic flat condition. Only in this way can we obtain the entropy
bound of the system. To some extent, this is coincident with two simple  calculations in \cite{Susskind:2005ss} (page 105 and 112) which show
that a large $r$ is needed to reach the entropy bound for photons. Those examples correspond concrete material system and are
included in our general calculation

Let us analyze the mathematical structure of the first law of holographic
thermodynamics. In this paper, we just care about the physics of isolated
systems. Consequently we consider the ADM mass $M$ and constant density $%
\rho $ are parameters rather than variables. With the only variable $r$, the
holographic thermodynamic equation is totally differential.\newline

\section{Discussion and Conclusion}

On Verlinde and Li's projects, gravity is treated as entropy force which is not related to elementary particles, while the
chemical potential is related to the variation of the number of particles in thermodynamics. In the first model, the screen
moving outward in the star leads to the increase of particles, therefore the holographic chemical potential appears in the screen, whereas the chemical potential in the second model does not arise owing to the screen moving in the vacuum. Since the specific matter constitution is not considered, the holographic chemical
potential cannot be deduced from the first principle in this paper. We can only judge the origin of the chemical potential from simple discussion.

Since the gravity is not fundamental and have no relevant elementary particles, we can relate it to phonon. The degrees of freedom are equal to the area in our convention. While the screen expanding in the vacuum, the degrees of freedom also rise without particles, thus it is contributed by the particles and gravity respectively. The effect of gravity in the holographic screen is the increasing number of modes of simple harmonic oscillation. A larger screen area implies the possibility of excitation of higher energy levels. We know the phonon have no chemical potential, therefore the screen moving in vacuum will not lead a chemical potential as the second case.

Having calculated these two special examples, we find that entropy bound is
a natural result of the first law of thermodynamics on the holographic
screen. However, more
research on holographic chemical potential need to be done in future.

\section{Acknowledgments}

We thank Miao Li and Xiao-Dong Li for valuable discussion and suggestion.

\end{CJK*}
\end{document}